\documentclass[aps,prb,onecolumn,superscriptaddress,10pt]{revtex4-2}
\pdfoutput=1 
\usepackage[utf8]{inputenc}
\usepackage[english]{babel}
\usepackage[T1]{fontenc}
\usepackage[pdftex]{graphicx}
\usepackage{amsmath}
\usepackage{cancel}
\usepackage{amsthm}
\usepackage{amssymb}
\usepackage{braket}
\usepackage{xcolor}
\usepackage{dcolumn}
\usepackage{bbm}
\usepackage{bbold}
\usepackage[normalem]{ulem}
\usepackage{footmisc}
\usepackage{xcolor}
\usepackage{booktabs}
\usepackage{comment}
\usepackage{multirow}
\usepackage[colorlinks=true,linkcolor=blue,citecolor=blue,urlcolor=blue,unicode]{hyperref}

\usepackage{amsmath}
\usepackage{bm}
\usepackage{graphicx}
\usepackage{subfigure}
\usepackage{ mathrsfs }
\usepackage{epstopdf}
\usepackage{ marvosym }
\usepackage{ tipa }
\usepackage{geometry}
\usepackage{ wasysym }
\usepackage{ulem}
\usepackage{indentfirst}
\usepackage{amsfonts,amssymb,dsfont}
\usepackage{accents}

\usepackage{setspace}
\usepackage{threeparttable}
\usepackage{amsmath,mathtools,amsthm}
\usepackage{array}
\usepackage{extarrows}
\usepackage{upgreek}
\usepackage{orcidlink}

\makeatletter
\renewcommand*\env@matrix[1][\arraystretch]{%
	\edef\arraystretch{#1}%
	\hskip -\arraycolsep
	\let\@ifnextchar\new@ifnextchar
	\array{*\c@MaxMatrixCols c}}
\makeatother

\begin{document}
	\title{Non-Markovian to Markovian Transition in a Driven Single-Site Holstein Model}
	\author{Chen-Huan Wu 
		\orcidlink{0000-0003-1020-5977} }
	\thanks{chenhuanwu1@gmail.com}
	\affiliation{College of Physics and Electronic Engineering, Northwest Normal University, Lanzhou 730070, China}

	\begin{abstract}

		We investigate the non-adiabatic dynamics of the Holstein mechanism in a single-site model driven by a time-dependent perturbation. 
		Unlike standard quenching protocols, we consider a unitary evolution where the driving field introduces a time-dependent displacement to the boson operators.
		The resulting dynamics reveal a dynamical quantum transition, evidenced by a distinct crossover from short-time non-Markovian behavior (characterized by power-law decay) to long-time Markovian-like behavior (exponential decay). 
		This transition is consistently manifested in the polaronic shift, bosonic energy, and the population dynamics of the reduced density matrix. 
		Furthermore, we analyze the emergence of effective non-unitary behavior. 
		We demonstrate that while the global evolution remains unitary, the emergence of a degenerate ground state manifold at long times acts as an effective environment. 
		Under a finite resolution threshold, the resulting fluctuations and mixing outweigh the coherent dynamics, leading to an effective unitary-to-nonunitary transition characterized by high mixedness and entanglement.

	\end{abstract}

\maketitle
\section{Introduction}

		The interaction between electrons and lattice vibrations (phonons) is a fundamental problem in condensed matter physics, leading to the formation of a quasi-particle known as the polaron, where an electron is "dressed" by a cloud of virtual phonons. The physical properties of these polarons, including their stability and phase transitions, are fundamentally influenced by whether the system is in an adiabatic or non-adiabatic regime \cite{Altman E}. 
		The physics of polarons is broadly separated into two limits. In adiabatic limit where the electron's Fermi velocity significantly exceeds the sound velocity, electron-phonon scattering is elastic, and interband transitions vanish, preventing electron energy loss \cite{Ragni,Johansson}.
		
		In this paper, we focus on the nonadiabatic regime, which becomes dominant under strong electron-phonon coupling. In this regime, a non-relativistic electron moving faster than the sound wave can interact with acoustic phonons, allowing it to lose energy via inelastic phenomena such as Cherenkov radiation\cite{Burin}. This regime is crucial for understanding complex polaron formation, as strong non-adiabaticity can lead to significant modifications in the energy spectrum, such as the avoided crossing in the polaron band structure \cite{Fetherolf,Alexandrov}. Such avoided crossings and related strong-interaction phenomena have also been observed in ultracold Fermi atomic systems \cite{Wu2,Pausch}.
		
		We employ the single-site Holstein model as a canonical framework for investigating these nonadiabatic, strong-coupling polaron dynamics. While this model captures the essential physics of an electron's self-trapping due to local interaction with bosons, our work introduces a novel perspective by studying the model under a time-dependent perturbation that drives the system's unitary evolution.
		Specifically this time-dependent perturbation breaks the degeneracies of the composite electron-boson Hilbert space. It introduces a time-dependent Hermitian part (driving term) to the boson operators. This Hermitian contribution leads the ground state of the annihilation boson operator to deviate from the classical coherent Gaussian state, driving the system away from equilibrium.
		
		The main goal of this work is to demonstrate that the resulting unitary dynamics reveal a quantum phase transition. We show that this transition manifests as a distinct crossover from non-Markovian behavior (characterized by power-law decay) at short times to Markovian behavior (exponential decay) at long times. 
		This signature is numerically verified in key observables, including the polaronic shift, the bosonic energy, and the dynamics of the reduced density matrix. 
		Unlike scenarios where decoherence arises from external bath effects or the orthogonality catastrophe in 1D systems \cite{Rosch A,Kantian A,Meden V}, our single-site model provides a clean environment to study the system's intrinsic non-Markovian to Markovian transition driven solely by the internal electron-phonon coupling under unitary evolution.

	\section{Model}
	
We consider the single-site Holstein model in the Schrödinger picture. The total time-dependent Hamiltonian is given by
\begin{equation} 
	\begin{aligned}
H(t) = H_{0} + H_{p}(t),
		\end{aligned}
\end{equation}
where the static part $H_{0}$ accounts for the electron, phonon, and their interaction
\begin{equation} 
	\begin{aligned}
H_{0} = \epsilon c^{\dagger}c +H_{b}+H_{eb}. 
		\end{aligned}
\end{equation}
Here, $c^{\dagger}$ ($c$) and $b^{\dagger}$ ($b$) are the time-independent creation (annihilation) operators for electrons and phonons, respectively.To investigate the non-equilibrium dynamics under non-adiabatic conditions, we introduce a time-dependent perturbation $H_{p}(t)$ that linearly drives the phonon field
\begin{equation} 
	\begin{aligned}
H_{p}(t) = f(t)(b^{\dagger} + b),
		\end{aligned}
\end{equation}
where $f(t)$ is a time-dependent external field (e.g., a laser pulse). 
Here we introduce a time-dependent driving field $f(t)$ that couples to the lattice displacement. The perturbation is given by $H_p(t) = f(t)(b^\dagger + b)$. Since $(b^\dagger + b) \propto x$ corresponds to the Hermitian position quadrature, this term introduces a time-dependent Hermitian part to the boson dynamics, driving the system away from its equilibrium vacuum state.
This term explicitly introduces the time-dependence into the system, breaking the time-translation symmetry and leading to the unitary evolution studied below.
$\dot{b}(t)=i[H(t),b]=-i\omega b(t)-ig\omega c^{\dag}c-if(t)$.

Due to the single-site consideration,
the bare bandwidth is irrelevant here.
Note that ${\rm Tr}[(c^{\dag}c)^2]={\rm Tr}[c^{\dag}c]^2/N$, precluding the Hubbrd-type interacion effect.
We consider the total Hamiltonian in a N-by-N Hilbert space,
and a time-dependent perturbation cause breaks the degeneracies of original composite Hilbert spaces.
This guarantees the unitary evolution of $H(t)$, 
which is distinct from the Heisenberg picture.
The bosonic term reads
\begin{equation} 
	\begin{aligned}
	H_{b}=\frac{p^{2}}{2}+\frac{1}{2}\omega^2 x^{2}
	=\omega(b^{\dag}b+\frac{{\bf I}}{2}),
		\end{aligned}
\end{equation}
where 
$b=\sqrt{\frac{\omega}{2}}(x+\frac{ip}{\omega}),\ 
b^{\dag}=\sqrt{\frac{\omega}{2}}(x-\frac{ip}{\omega}),\ 
x=\sqrt{\frac{1}{2\omega}}(b+b^{\dag}),\ 
p=-i\sqrt{\frac{\omega}{2}}(b-b^{\dag})$.
The boson number operator $b^{\dag}b$ is $N$-by-$N$ Hermitian matrices with a highly degenerate spectrum 
and are completely degenerate initially (at t=0),
reflecting the internal degrees of freedom. Also, we consider the case that the fermion number operator is always at an equilibrium steady state throughout the evolution to eliminate the potential effect of its fluctuation on the polaron shift. 
For electron-phonon interaction, we consider the Holstein mechanism
\begin{equation} 
	\begin{aligned}
	H_{eb}=g\omega c^{\dag}c(b^{\dag}+b)
	=g\sqrt{2\omega}\omega c^{\dag}cx.
		\end{aligned}
\end{equation}

\section{Results}
\subsection{Polaronic shift}
Different to the classical harmonic oscillator where the time-dependence of boson operator can be
extracted as a phase factor, which is necessary for the Lang-Firsov transformation,
the boson operator here cannot.
Instead,
we can separate the above boson operators (non-Hermitian) 
into the quantum fluctuation part and classical displacement part, such that 
\begin{equation} 
	\begin{aligned}
	&b(t)= b(0)e^{-i\omega t} - i \int_0^t d\tau (g\omega c^{\dag}c+f(\tau)) e^{-i\omega(t-\tau)}
= b_{0}(t) + \beta(t) - g c^\dagger c (1 - e^{-i\omega t}),\\
	&b^{\dagger}(t) = b^{\dagger}(0)e^{i\omega t} + i \int_0^t d\tau (g\omega c^{\dag}c+f(\tau)) e^{i\omega(t-\tau)}
	= b_{0}^{\dagger}(t) + \beta^*(t)-g c^\dagger c (1 - e^{i\omega t}) ,\\
&x(t)=x_{1}(t)+x_{2}(t)+x_{3}(t):=
 \sqrt{\frac{1}{2\omega}} \left( b(0)e^{-i\omega t} + b^{\dagger}(0)e^{i\omega t} \right)
 +\sqrt{\frac{1}{2\omega}} \left( \beta(t) + \beta^*(t) \right)
-g c^\dagger c \sqrt{\frac{2}{\omega}} (1 - \cos(\omega t)),\\
&p(t)=p_{1}(t)+p_{2}(t)+p_{3}(t):=
 -i\sqrt{\frac{\omega}{2}} \left( b(0)e^{-i\omega t} - b^{\dagger}(0)e^{i\omega t} \right)
 -i\sqrt{\frac{\omega}{2}} \left( \beta(t) - \beta^*(t) \right)
 -\sqrt{2\omega}gc^{\dag}c{\rm sin}(\omega t),
	\end{aligned}
\end{equation}
where we use 
$-i g\omega c^\dagger c \int_0^t d\tau e^{-i\omega(t-\tau)} = -ig\omega c^\dagger c \left[ \frac{e^{-i\omega(t-\tau)}}{i\omega} \right]_0^t = -g c^\dagger c (1 - e^{-i\omega t})$,
$i g\omega c^\dagger c \int_0^t d\tau e^{i\omega(t-\tau)} = -g c^\dagger c (1 - e^{i\omega t})$.
The commutation relation is satisfied:
$[\beta(t),\beta^{*}(t)]=[- g c^\dagger c (1 - e^{-i\omega t}), -g c^\dagger c (1 - e^{i\omega t})] =0$ and $[b_{0}(t),b^{\dag}_{0}(t)]=1$.
Under the time-dependent perturbation $H_p(t) = f(t)(b^\dagger+b)$, the Heisenberg operators evolve as the sum of a free quantum evolution and a classical displacement and a polaron coupling-induced displacementy where $b_0(t) = b(0)e^{-i\omega t}$ is the free evolution operator and $\beta(t) = -i \int_0^t d\tau f(\tau) e^{-i\omega(t-\tau)}$ is the perturbation-induced displacement. 
The quantum Fluctuations ($x_1$ and $p_{1}$) originate from the free evolution $b(0)$, these terms represent the intrinsic quantum uncertainty. They satisfy ${\rm Tr}[x_{1}]=0$ but ${\rm Tr}[x_{1}^2] \neq 0$, preserving the commutation relations.
The classical mean-fields ($x_2$, $p_{2}$ and $x_{3}$, $p_{3}$) arise from the external drive $f(t)$ and the electron-phonon coupling $g$, respectively. 
Within the single-electron subspace, these are proportional to the identity matrix (c-numbers). They satisfy the scalar trace relation ${\rm Tr}[O^2] = ({\rm Tr}[O])^2/N$, representing well-defined classical trajectories in phase space.

Thus the position and momentum operators separate into quantum fluctuation parts ($x_{1}$ and $p_{1}$,), classical mean-field parts ($x_{2}$ and $p_{2}$) where $x_{2}(t) = \sqrt{\frac{2}{\omega}}\text{Re}[\beta(t)]$ and $p_{2}(t) = \sqrt{2\omega}\text{Im}[\beta(t)]$ are the time-dependent classical expectations driven by the external field, and the polaornic shift
$x_{3}(t)$ and $p_{3}$.
Note that all these components $x_{i},p_{i}$ ($i=1,2,3$) are Hermitian matrices.
$x_2(t)$ is the unique term responsible for the relaxation dynamics (decay), whereas $x_1(t)$ and $x_3(t)$ represent persistent quantum fluctuations and coherent polaronic oscillations, respectively, which do not decay with time.

The classical parts $x_{2}(t)$ and $p_{2}(t)$ are proportional to the identity matrix in the boson subspace (c-numbers derived from the driving field), satisfying the relations ${\rm Tr}[x_{2}^2] = ({\rm Tr}[x_{2}])^2/N$ and  ${\rm Tr}[p_{2}^2] = ({\rm Tr}[p_{2}])^2/N$. In contrast, the quantum parts $x_{1}(t)$ and $p_{1}(t)$ describe the fluctuations around the classical trajectory and do not satisfy this trace relation.
We consider the case where the phase-space distribution remains a product of Gaussian Wigner distributions. The minimum uncertainty condition $\langle x_{1}^2\rangle = (4\langle p_{1}^2\rangle)^{-1}$ holds for the quantum fluctuation part ($\langle x_1^2 \rangle = \frac{1}{2\omega}$, $\langle p_1^2 \rangle = \frac{\omega}{2}$), consistent with the fact that the reduced density matrix describes a pure coherent state (displaced vacuum) throughout the unitary evolution.To decouple the electron-phonon interaction, we apply the standard Lang-Firsov (polaron) transformation, defined by the unitary operator $ e^S$, with the anti-Hermitian generator
\begin{equation}
	S = g c^{\dagger}c (b^{\dagger} - b) = -i g\sqrt{2\omega} c^{\dagger}c p_{1}(0).
\end{equation}
Note that we define $S$ using the time-independent Schrödinger operators to establish a consistent polaron frame. Applying this transformation $\overline{O} = e^S O e^{-S}$ to the total Hamiltonian $H(t)$, we obtain the transformed Hamiltonian
\begin{equation}
\overline{H}_0 = \epsilon c^\dagger c + \omega (b^\dagger - g c^\dagger c)(b - g c^\dagger c) + g\omega c^\dagger c (b^\dagger - g c^\dagger c + b - g c^\dagger c)
= (\epsilon - g^2\omega) c^\dagger c + \omega b^\dagger b+\frac{\omega}{2},
\end{equation}
where we use 
$(c^\dagger c )^2=c^\dagger c $,
$e^S c^\dagger c e^{-S} = c^\dagger c$,
$e^S b e^{-S} = b - [b, S] = b - g c^\dagger c$,
$e^S b^\dagger e^{-S} = b^\dagger - g c^\dagger c$.
The transformed driving term becomes
\begin{equation} 
	\begin{aligned}
\overline{H}_{p}(t) 
= f(t) [ (b^\dagger - g c^\dagger c) + (b - g c^\dagger c) ]
= f(t)(b^\dagger + b) - 2g f(t) c^\dagger c.
	\end{aligned}
\end{equation}
For self-energy $\Sigma(t)$ arising from the external drive, it reads $\Sigma(t) = -g^2\omega - 2gf(t)$. Thus the external perturbation acts as a time-varying potential on the dressed electron, modulating its binding energy $-g^2\omega$.

Due to the unitary evolution generated by the time-dependent drive, the expectation value of the boson Hamiltonian does not correspond to a simple thermal decay but follows the dynamics of the displacement field.
Next we analyze the polaronic shift.
Applying the Lang-Firsov transformation, the effective electron-phonon coupling term (related to the self-energy) becomes
\begin{equation} 
	\begin{aligned}
\frac{1}{g\omega N}\langle \Sigma_{eb}(t) \rangle
&= \frac{1}{N} \langle e^{S}(b^{\dagger}(t)+b(t))e^{-S} \rangle \\
&= \frac{1}{N} \langle \left[ b^\dagger(0)e^{i\omega t} + b(0)e^{-i\omega t} \right] + [\beta^*(t) + \beta(t)] - g c^\dagger c - g c^\dagger c  \rangle \\
&= \frac{1}{N} \left( (\beta^*(t) + \beta(t)) - 2g \right)
	\end{aligned}
\end{equation}
where we use
\begin{equation} 
	\begin{aligned}
	e^S b(t) e^{-S} &= e^S \left[ b(0)e^{-i\omega t} + \beta(t) - g c^\dagger c + g c^\dagger c e^{-i\omega t} \right] e^{-S} \\
	&= \left( e^S b(0) e^{-S} \right) e^{-i\omega t} + \beta(t) - g c^\dagger c + g c^\dagger c e^{-i\omega t} \\
	&= (b(0) - g c^\dagger c) e^{-i\omega t} + \beta(t) - g c^\dagger c + g c^\dagger c e^{-i\omega t}\\
		&= b(0) e^{-i\omega t} + \beta(t) - g c^\dagger c ,\\
	e^S b^{\dag}(t) e^{-S} & = b^\dagger(0)e^{i\omega t} + \beta^*(t) - g c^\dagger c,
	\end{aligned}
\end{equation}
Note that
$\langle b_0^\dagger(t) + b_0(t) \rangle=
\langle  b^\dagger(0)e^{i\omega t} + b(0)e^{-i\omega t} \rangle = 0$,
$e^{S} b_0(t) e^{-S} = e^{S} (b(0)e^{-i\omega t}) e^{-S} = \left( e^{S} b(0) e^{-S} \right) e^{-i\omega t}= \left( b(0) - gc^\dagger c \right) e^{-i\omega t}$
and thus $e^{S} (b_0(t)+b_{0}^{\dag}(t)) e^{-S} 
=\left( b^\dagger(0)e^{i\omega t} + b(0)e^{-i\omega t} \right) - gc^\dagger c (e^{-i\omega t}+e^{i\omega t})
= \left( b^\dagger(0)e^{i\omega t} + b(0)e^{-i\omega t} \right) - 2gc^\dagger c \cos(\omega t)$.
The transformed term involving the quantum fluctuations, $e^{S}(b_0^{\dagger}(t)+b_0(t))e^{-S}$, contains operators oscillating at frequency $\omega$. However, since the system remains in the vacuum state with respect to the fluctuation operators $b_0$, their expectation value vanishes, $\langle b_0^{\dagger}(t)+b_0(t) \rangle = 0$. The non-trivial time-dependence of the polaronic shift $\Sigma(t)$ is entirely dominated by the classical displacement part $\beta(t)$, which yields the observed transition from power-law to exponential decay.

Note that the quantum fluctuation terms vanish under the expectation value in the vacuum state, i.e., $\langle b_0(t) \rangle = 0$. The dynamics are thus strictly determined by the classical displacement $\beta(t)$.
Under the driving protocol $f(t)$, the asymptotic behavior of the displacement is,
\begin{equation} 
	\beta^*(t) + \beta(t) \sim \left\{
	\begin{array}{ll}
		(2-\ln\omega)\frac{\omega}{t}, & (t \ll 1) \\
		(2-\ln\omega) e^{-t/\omega}, & (t \gg 1)
	\end{array}
	\right.
\end{equation}
Fig.\ref{2} shows 
the time evolution of the classical component of the position operator $x_2(t)$. The solid lines represent the numerical results, while the orange and green dashed lines indicate the fitting for the early-stage power-law decay and late-stage exponential decay, respectively. 
This confirms that the macroscopic displacement dynamics are driven by the external perturbation.

By inverting the equation of motion $\dot{\beta} = -i\omega\beta - if(t)$, we can deduce that the observed transition in the displacement $\beta(t)$ directly reflects the spectral properties of the external perturbation $f(t)$. A driving field that transitions from a power-law decay ($f(t) \sim t^{-\alpha}$) to an exponential decay ($f(t) \sim e^{-\gamma t}$) will naturally induce the non-Markovian to Markovian crossover observed in Fig.2.
From the relation $f(t) = i\dot{\beta}(t) - \omega\beta(t)$, the observed crossover in $\beta(t)$ from power-law ($f(t) \approx i\dot{\beta}(t)\approx i(-\alpha t^{-\alpha-1}) - \omega t^{-\alpha} \approx -\omega t^{-\alpha}$) to exponential ($f(t)\approx i\dot{\beta}(t) \approx i(-\gamma e^{-\gamma t}) - \omega e^{-\gamma t} = -(\omega + i\gamma) e^{-\gamma t}$) decay implies a similar transition in the temporal correlations of the perturbation $f(t)$.
This confirms that our time-dependent perturbation effectively mimics a bath interaction that transitions from a noise regime of non-Markovian with long-range memory to a white noise regime (Markovian with short-range memory). This provides a dynamical origin for the observed phase transition in the polaron formation process.

\begin{figure}
		\includegraphics[width=0.6\linewidth]{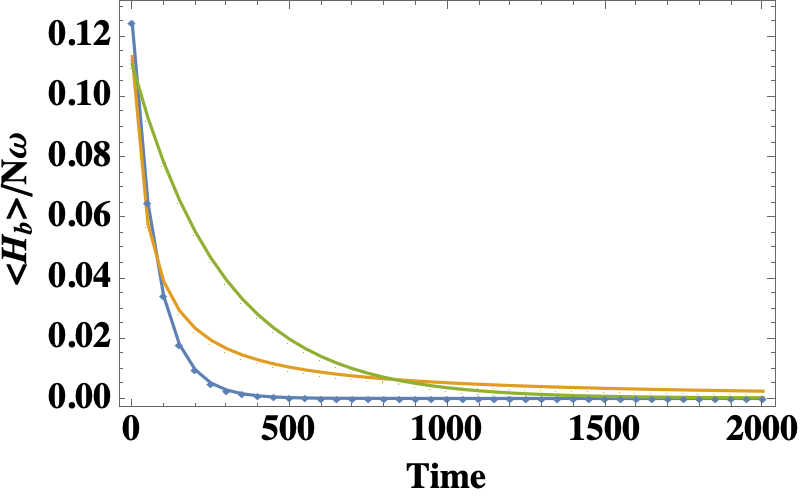}
		\caption{Expectation of boson operator on the single electron basis state.
		at long-time.
}
		\label{M61}
\end{figure}

\begin{figure}
		\includegraphics[width=0.6\linewidth]{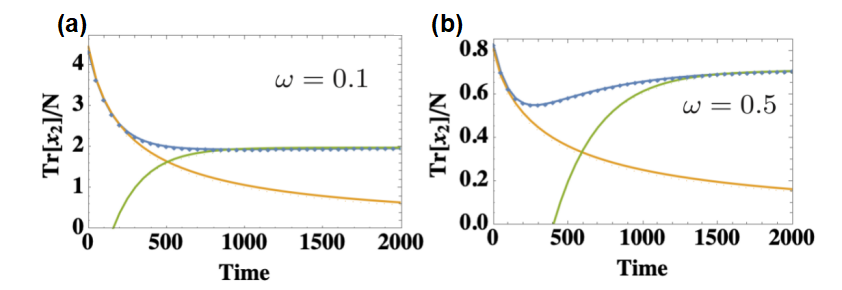}
		\caption{Trace of the classical part of position operator $x_{2}(t)$.
The orange and green lines fit the early and latter stages using power-law and exponential decay, respectively.}
\label{2}
\end{figure}

This transition implies a crossover from power-law decay to exponential decay.
The total boson energy expectation value behaves as:
\begin{equation} 
	\frac{1}{\omega}\langle H_{b}(t) \rangle = \langle b^\dagger(t)b(t) \rangle + \frac{1}{2} = |\beta(t)|^2 + \frac{1}{2}
	\sim \left\{
	\begin{array}{ll}
		\frac{1}{t}, & (t \ll 1) \\
		e^{-t}, & (t \gg 1)
	\end{array}
	\right.
\end{equation}
where $\langle b^\dagger(t) b(t) \rangle = \langle (b_0^\dagger(t) + \beta^*(t)) (b_0(t) + \beta(t)) \rangle= \langle b_0^\dagger b_0 \rangle + \beta(t)\langle b_0^\dagger \rangle + \beta^*(t)\langle b_0 \rangle + |\beta(t)|^2= |\beta(t)|^2$,
as shown in Fig.\ref{M61}.
Consequently, the particle number behaves as $\langle b^{\dagger}(t)b(t) \rangle \sim \frac{1}{t+1/N}$ for small $t$, and $\sim 1+e^{-t}$ for large $t$.
This transition from power-law to exponential decay signifies that the system's relaxation dynamics shift from a non-Markovian regime (memory effects) to a Markovian-like regime at long times, governed by the time-dependence of the external drive $f(t)$.

\begin{figure}
		\includegraphics[width=0.6\linewidth]{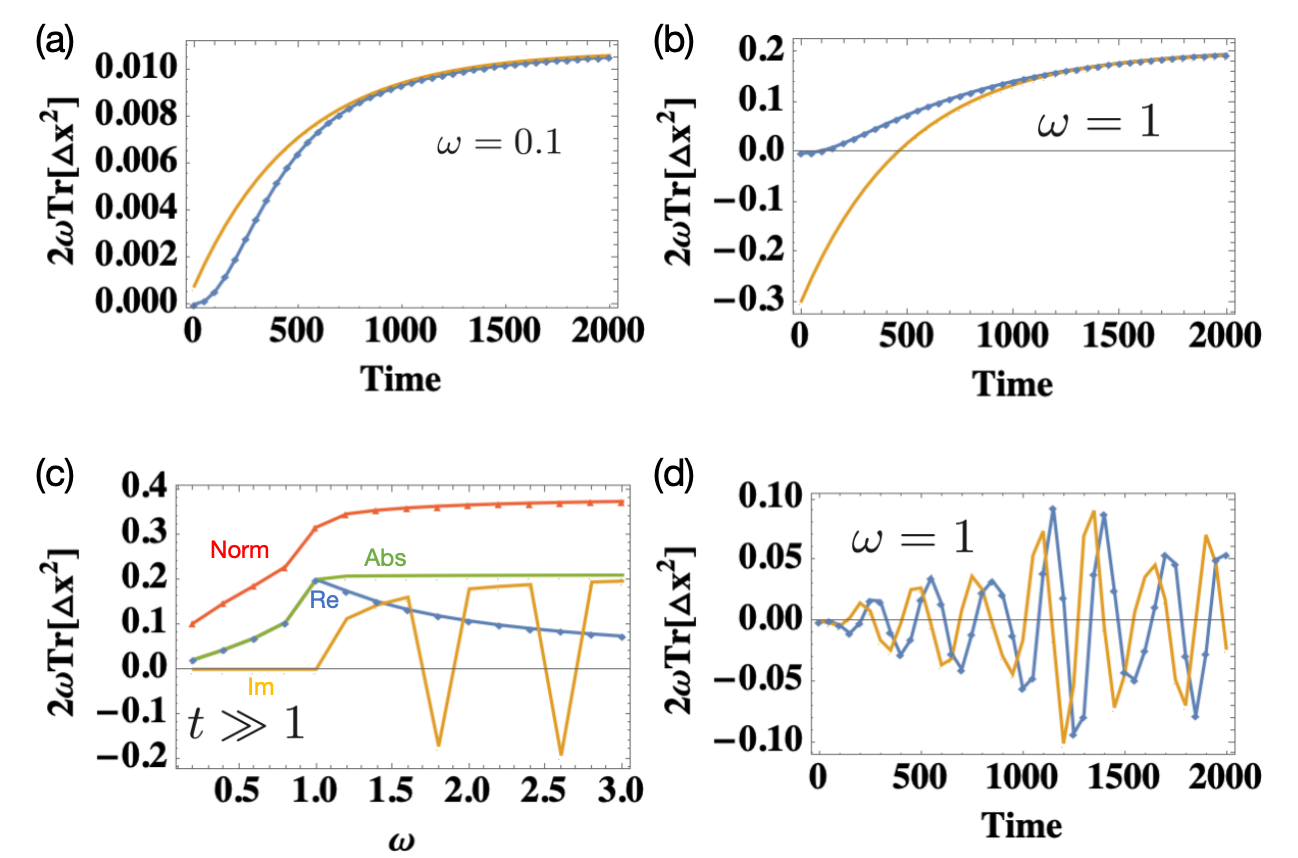}
		\caption{Expectations of $\Delta x^2$ as a function of time ((a)-(b)) and frequency ((c)). 
Above the critical frequency $\omega_{c}=1$, the real part star to decline and nonzero imaginary part emerges.
(d)The same with (a)-(b) but replacing the $S$ by $S'$.}
		\label{3}
\end{figure}

\subsection{Effect of transformation on the boson operators}

Since the generator of the transformation $S$ involves the boson operators, it does not commute with the position operator unless the coupling vanishes. We define the shift in the position operator induced by the transformation as
\begin{equation}
	\sqrt{2\omega}\Delta x(t):=e^{S}(b^{\dag}(t)+b(t))e^{-S}-(b^{\dag}(t)+b(t)).
\end{equation}
This operator difference arises from the non-commutativity between the generator $S$ and the field operators. 
Since the transformation is unitary, the trace of the squared position operator must be invariant, i.e., ${\rm Tr}[(e^S x(t) e^{-S})^2] = {\rm Tr}[x(t)^2]$. 
Substituting $e^S (b^\dagger + b) e^{-S} = (b^\dagger + b) + \sqrt{2\omega}\Delta x(t)$ into this invariance condition, we derive the following identity
\begin{equation} \label{eq:trace_identity}
	\begin{aligned}
		{\rm Tr}[2\omega \Delta x(t) ^2] + 2{\rm Tr}[(b^{\dag}(t)+b(t))\sqrt{2\omega} \Delta x(t)]=0.
	\end{aligned}
\end{equation}
This relation allows us to quantify the magnitude of the polaronic shift dynamically.
For the regime $\omega\le 1$, the trace of the squared shift exhibits distinct asymptotic behaviors
\begin{equation} 
	\begin{aligned}
		{\rm Tr}[2\omega \Delta x(t) ^2] 	
		\sim \left\{
		\begin{array}{ll}
			0, & (t \ll 1)\\
			\omega(1-e^{-t}), & (t \gg 1)
		\end{array}
		\right.
	\end{aligned}
\end{equation}
which becomes linear in $\omega$ at long times.
The exact numerical result of ${\rm Tr}[2\omega \Delta x(t) ^2]$ is shown in Fig.\ref{3}.
An approximate analytical form is also provided in the Appendix.A.

In Holstein model and driven harmonic oscillators, $\omega=1$ here represents a resonance point where the phonon frequency matches the electronic or driving energy scale. 
The expectation value of a Hermitian observable like position squared ($\Delta x^2$) must be real. 
The emergence of an imaginary part for $\omega > 1$ implies that the polaron ansatz (the transformation $S$) breaks down or becomes unstable in high-frequency regime.
This indicates a phase transition. Below $\omega=1$, a stable polaron forms (real displacement). Above $\omega=1$, the system enters a delocalized or radiative regime where the static polaron picture fails and the complex values indicating the decay or instability of that specific bound state solution.

Fluctuations signifying the non-local correlations can be observed in Fig.\ref{3}(d). Here we consider a modified transformation generator $S'=c^{\dag}(0)c(t)(b^{\dag}(t)-b(t))$. 
Unlike the static $S$, this operator $S'$ is no longer strictly anti-Hermitian for $t \neq 0$ due to the time-dependence of the electron operators in the Heisenberg picture.
Consequently the nonzero value of ${\rm Tr}[2\omega \Delta x(t)^2]$ in this context implies a deformation of the phase space distribution, analogous to the squeezing effect generated by canonical transformations of the form $e^{r(b^2-(b^{\dag})^2)}$,
as a time-dependent transformation often induces Rabi-like oscillations or squeezing effects (periodic expansion and contraction of phase space), which looks exactly like the ringing shown here.
The rapid oscillations seen in Fig.\ref{3}(d) are the signature of non-local correlations and squeezing or breathing modes in the boson field. Unlike the smooth saturation of the standard shift (Fig.\ref{3}(a)), squeezing induces time-dependent variances that oscillate at $2\omega$.
Thus the nonzero value of $\text{Tr}[2\omega \Delta x(t)^2]$ implies a deformation of the phase space distribution. This is analogous to the squeezing effect, where the vacuum fluctuation circle is distorted into an ellipse. As this deformed state undergoes free evolution (rotation in phase space at frequency $\omega$), its projection onto the position axis (variance) expands and contracts twice per revolution. 
This generates the characteristic oscillations at $2\omega$ (twice the natural phonon frequency) observed in Fig.3(d).

\section{Dynamics of reduced density matrix}

\begin{figure}
		\includegraphics[width=0.6\linewidth]{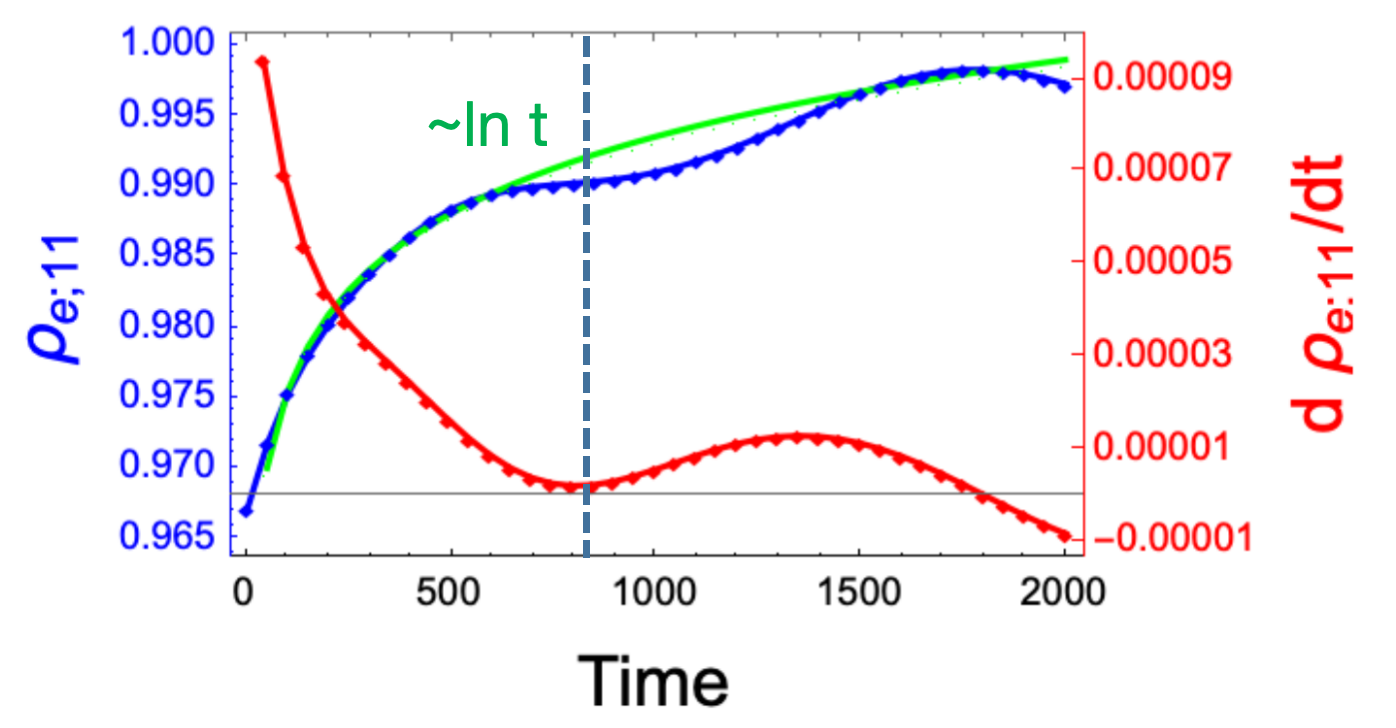}
		\caption{First diagonal element of the reduced density matrix $\rho_{e}$
		and its time derivative.
		The vertical dashed line indicate the time where
		 the degenerated ground states of $H$ emege.}
		\label{redu}
\end{figure}

We next focus on the dynamics of reduced density matrix $\rho_{e}={\rm Tr}_{b}\rho_{tot}$
(partial trace over the bosonic bases)
where $\rho_{tot}=|\psi_{0}\rangle\langle\psi_{0}|$ with $|\psi_{0}\rangle$ the eigenstate of $H$ corresponding to lowest
eigenvalue,
i.e., $\rho_{tot}=\sum_{ij}c_{i}c_{j}^{*}|\Psi_{i}\rangle\langle\Psi_{j}|$ where $c_{i}$ is the $i$-th component of 
$|\psi_{0}\rangle$, and 
$|\Psi_{i}\rangle$ is the $i$-th state of the combined basis.
Note that the rank of $H$ is lowered with increasing time, in this case we use 
$\rho_{tot}=\frac{1}{d}\sum_{j=1}^{d} |\psi_{0;j}\rangle\langle\psi_{0;j}|$ with $d$ the ground state degeneracy,
in which case it requires further orthonormalized to keeping ${\rm Tr}\rho_{e}=1$.
While for $\rho_{tot}$ in mixed state which corresponds to the case of finite temperature, 
we have $\rho_{tot}=\sum_{j}e^{-\beta E_{j}}|\psi_{j}\rangle\langle\psi_{j}| /\sum_{j}e^{-\beta E_{j}}$.
But we stick to the zero temperature case in this paper,
thus we do not adopt this mixed state.
The resulting reduced electron density $\rho_{e}$ is a 2-by-2 Hermitian matrix,
with the two diagonal elements describe the populations of two states in electron Hilbert space,
and the two off-diagonal elements describe the coherence between the two states.
The result shows that, at short-time (far before the emergence of degenerated ground state of $H$),
$\rho_{e;11}\sim {\rm ln}t (=1-\rho_{e;22})$,
and reasonably, $\frac{d\rho_{e;11}}{dt}\sim \frac{1}{t} (=-\frac{d\rho_{e;22}}{dt})$,
as shown in Fig.\ref{redu}.
The logarithmic increase in population $\rho_{e;11}$ reflects a extremely slow dynamics,
and the system is not reaching a steady state.
The power-law decay in $\frac{d\rho_{e;11}}{dt}$ again signifying the Non-Markovian Memory effect in early stage during the evolution.
While exponential decay emerges at longer time (but before the emergence of degenerated ground state of $H$),
$\rho_{e;11}\sim -e^{-t}$ and $\frac{d\rho_{e;11}}{dt}\sim e^{-t}$.

\begin{figure}
		\includegraphics[width=0.8\linewidth]{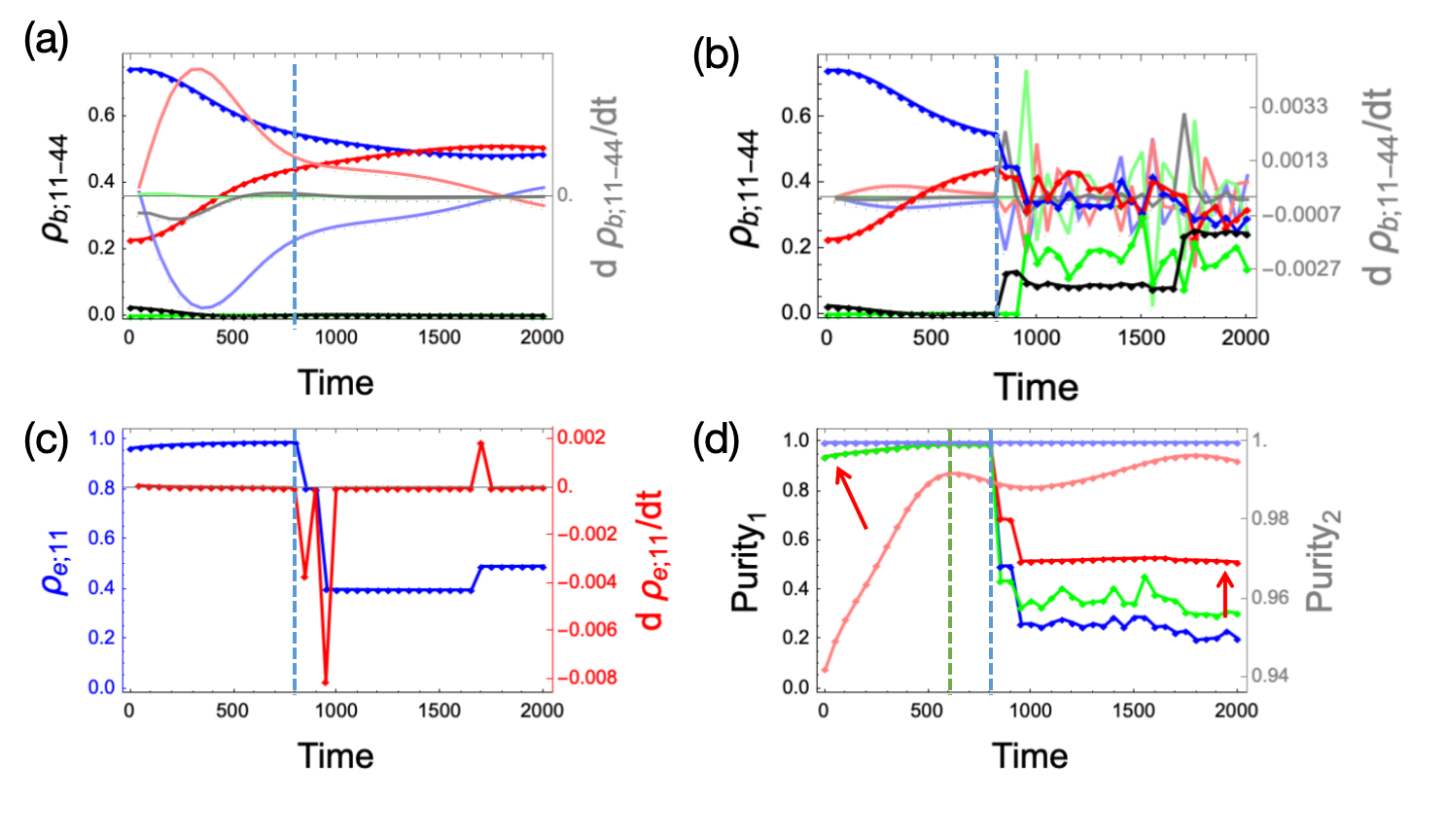}
		\caption{(a) The four diagonal elements (populations) of $\rho_{b}$ 
		for unitary evolution and pure total density, obtained by setting an infinitesimally small threshold value for judging the degenerate ground state.
(b)-(c) Populations of $\rho_{b}$ and $\rho_{e}$ when a finite threshold value is used. The non-unitary evolution and mixed total density appear at a critical time $t_{c}\approx 800$, as indicated by the vertical cyan dashed line.
(d) Purities of $\rho_{tot}$ (blue), $\rho_{e}$ (red), and $\rho_{b}$ (green) 
for finite threshold value (left axis) and infinitesimally  threshold value (right axis). 
The green dashed line indicates when the population starts to deviate from its logarithmic increase, which signifies the emergence of a Markovian process. The left red arrow indicates the lowered purity due to entanglement between subsystems, while the right red arrow indicates the lowered purity due to entanglement between the subsystem and the bath. Obviously, the effect of the latter dominates the former. That is why non-Markovian type fluctuations outweigh the Markovian dynamics in the late stage.}
\label{f7}
\end{figure}

As shown in Fig.\ref{redu}, 
$\rho_{e;11}$ is quite large and increase with time, 
thus $\rho_{e}$ and $\rho_{b}$ are close to purity state and the purity increase with time.
This is consistent with the increased purity of each subsystem as shown in Fig.\ref{f7}(d)
and the continously decrease of internal entanglement (negativity) of $\rho_{tot}$ as shown in Fig.\ref{nega}(b).
Thus
the nonMarkovian-to-Markovian transition can be well observed at least before the critical time $t_{c}$
of the emergent degenerated ground state of $H$,
during which the evolution is unitary and the system is nearly closed such that as the
total density getting thermalized by the environment with lowered internal entanglement, 
the subsystems have increased purity until the entanglement with the environment sets in 
(the begining of nonunitary evolution).
In other word,
in early stage, the weak nonMarkovianity leads to power-law monotonous decay;
while in late stage, the degenerate ground state of $H$ cause additional strong nonMarkovianity 
with significant
coherent exchange of information between system and bath,
which outweights the Markovian thermal process despite the 
decoherence dominates at the long-time and implying the enhanced entanglement.
Note that the Markovian dynamics in the late stage can still be observed in term of the observables regardless
of the density operators.
Thus, the perturbation, which is initially well-defined and coherent, 
in the late stage becomes random and is composed of 
many unobserved degrees of freedom due to the degeneracy of the system's ground state.
From Fig.\ref{f7}(d),
we also notice that $\rho_{b}$ and $\rho_{tot}$ have richer information exchange, in the late stage, compares to the $\rho_{e}$, as manifested in stronger fluctuation of purity,
which implies that, even for a separable multi-qubit system (see Fig.\ref{nega}(b)), the remaining classical correlations
can still affect the exchange of information with the environment.

The dissipative phase transition is only observable in the early stages of evolution. During this period, the dissipation-induced gain and loss can be ignored, and the total Hamiltonian maintains unitary evolution.
When the ground state is resolved with infinite precision (infinitesimal threshold), the total density matrix $\rho_{tot}$ remains a pure state $|\psi(t)\rangle\langle\psi(t)|$ governed by the Schrödinger equation. In this ideal limit, the system is closed, and entropy is generated solely in the subsystems ($\rho_e, \rho_b$) via entanglement between the electron and the phonon bath.

When we introduce a finite threshold value to define the ground state manifold, a transition occurs at a critical time $t_c$ where the ground state becomes highly degenerate.
For $t > t_c$, the degeneracy introduces an effective disorder. The evolution is no longer well-described by a single trajectory but requires a statistical mixture over the degenerate manifold, $\rho_{tot}=\frac{1}{d}\sum_{j=1}^{d} |\psi_{0;j}\rangle\langle\psi_{0;j}|$.
This coarse-graining acts as an information loss mechanism. As noted in Appendix B, this is equivalent to considering the perturbation as inherently random, which induces dephasing and mixing.
Thus
by setting a finite threshold value for the ground state, Fig.\ref{f7} presents the scenario where the total density matrix is considered to be in a mixed state. Once time exceeds a critical value, the total system's density matrix becomes mixed, implying that the system is indeed an open system and its evolution becomes non-unitary at long times. Here, strong dissipation drives the system toward a state akin to a highly fluctuating thermal state in a closed system (see Fig.\ref{f7}(b)).
This observation is consistent with the opening of the dissipative gap in the later stage of evolution. At this point, quantum fluctuations (under the zero-temperature condition) make continuous symmetry breaking and the formation of an ordered phase less likely.
Consequently, in finite-threshold regime, the total system effectively behaves as an open system. The "environment" here is identified with the unobserved internal degrees of freedom within the degenerate ground state manifold and the stochastic fluctuations of the perturbation.
 Strong dissipation drives the system toward a state akin to a highly fluctuating thermal state (see Fig.\ref{f7}(b)).
This leads to the unitary-to-nonunitary transition observed in Fig.\ref{nega}(b), where the total system entropy increases (purity decreases) after $t_c$,
due to the entanglement with the environment where the environment is the effective bath created by the disorder and degeneracy.
This observation is also consistent with the opening of a dissipative gap in the later stage of evolution, where the quantum fluctuations (under the zero-temperature condition) suppress continuous symmetry breaking, making the formation of an ordered phase less likely.

\section{Conclusion}

We investigated electron-phonon coupling in a single-site system characterized by unitary evolution intrinsic to the boson operators within the nonadiabatic limit. 
We focused on the regime where the phase-space distribution retains its Gaussian nature. Specifically, the variances of position and momentum satisfy the minimum uncertainty relation $\langle x^2\rangle=(4\langle p^2\rangle)^{-1}$, characteristic of a coherent state (or displaced vacuum) at zero temperature.
This minimum uncertainty is guaranteed by the fact that the reduced density matrices remain nearly pure states during the evolution, i.e., ${\rm Tr}[\rho_{e}^2]={\rm Tr}[\rho_{b}^2]\approx 1$. 
Since the total evolution is unitary and the global state remains pure (${\rm Tr}\rho_{tot}^2=1$), this high purity implies weak entanglement in the bipartite system, although entanglement is slightly enhanced at long times due to the emergence of Markovian dynamics and decoherence processes.

This observed transition from non-Markovian to Markovian dynamics, accompanied by the crossing from a zero dissipative gap to an open dissipative gap, implies a change in the scaling of correlations. This behavior is analogous to the transition from an area law to volume law scaling of entanglement, or the transition from higher to lower spatial dimensions with an enhanced role of dissipation.
We also note that this transition can be viewed as continuous in time: shifting from a time-dependent Lindbladian (governing the non-Markovian power-law decay stage) to a time-independent Lindbladian (governing the Markovian exponential decay stage), provided the total density matrix describes a pure state.
The evolution of the reduced state, formally written as $|\rho_{e}(t)\rangle\rangle=\mathcal{T} e^{\int^{t}_{0} \mathcal{L}(\tau)d\tau} |\rho_{e}(0)\rangle\rangle$, is analogous to continuous connected gapped Liouvillians in systems of finite spatial dimension mediated by long-range interactions\cite{Zhang}.

The unitary dynamics preserves the structure of the phase-space distribution during the Markovian process, characterized by slow relaxation where the electron-phonon interaction dominates over the influence of the phonon bath.
In this nonadiabatic limit, the simple Gaussian wave function approximation fails to fully capture the bosons, which can no longer be treated purely as a classical field, and the phononic time scale becomes comparable to or faster than the electronic time scale. Thus standard mean-field theory also fails in this regime.

Our results provide evidence of a dynamical phase transition, evidenced by distinct behaviors at short-time and long-time stages. The unitary dynamics and the boson-frequency dependence lead to a crossover from power-law decay to exponential decay, as observed in the polaronic shift, boson energy, and the dynamics of the reduced density matrix $\rho_{e}$.
Moreover, the effective mass renormalization $m^*/m(=e^{\Sigma(t)/\omega})$ is expected to follow the same scaling rule.
The gradual opening of the dissipative gap can be further verified by estimating the spectrum of the effective Lindbladian. 
This estimation, achievable through Lanczos iteration or quantum trajectory methods—as well as the investigation of boson squeezing ($(b^{\dag})^2+b^2$) and deviations from the minimum uncertainty condition driven by entanglement—will be the scope of our future work.

\renewcommand\refname{References}

\section{Appendix.A: Approximation form of ${\rm Tr}[2\omega \Delta x(t) ^2]$}

Using the following relations
\begin{equation} 
	\begin{aligned}
&{\rm Tr}[(e^{S}(b^{\dag}(t)+b(t))e^{-S})^2]={\rm Tr}[(b^{\dag}(t)+b(t))^2]
={\rm Tr}[e^{S}(b^{\dag}(t)+b(t))^2 e^{-S}],\\
&{\rm Tr}[e^{S}(b^{\dag}(t)+b(t))e^{-S}(b^{\dag}(t)+b(t))]={\rm Tr}[(b^{\dag}(t)+b(t))e^{S}(b^{\dag}(t)+b(t))e^{-S}],\\
&
{\rm Tr}[e^{S}(b^{\dag}(t)+b(t)) e^{S}(b^{\dag}(t)+b(t))e^{-S}]
={\rm Tr}[(b^{\dag}(t)+b(t))^2 e^{S}]={\rm Tr}[(b^{\dag}(t)+b(t))^2 e^{-S}]^*\\
&={\rm Tr}[e^{S}(b^{\dag}(t)+b(t))^2 ]={\rm Tr}[e^{-S}(b^{\dag}(t)+b(t))^2]^*
    			\end{aligned}
\end{equation}
we can obtain
\begin{equation} 
	\begin{aligned}
&{\rm Tr}[e^{S}(b^{\dag}(t)+b(t))^2 e^{-S}]-{\rm Tr}[e^{S}(b^{\dag}(t)+b(t))e^{-S}(b^{\dag}(t)+b(t))]
={\rm Tr}[e^{S}(b^{\dag}(t)+b(t)) [(b^{\dag}(t)+b(t)),e^{-S}]]\\
&={\rm Tr}[e^{S}(b^{\dag}(t)+b(t)) (-[(b^{\dag}(t)+b(t)),S]
+\sum_{n=2} \frac{(-1)^n}{n!}[(b^{\dag}(t)+b(t)),S^n])]\\
&= {\rm Tr}[e^{S}(b^{\dag}(t)+b(t)) (e^{S}(b^{\dag}(t)+b(t))e^{-S}-(b^{\dag}(t)+b(t))
-\sum_{n=2}\frac{1}{n!}[S^{(n)},(b^{\dag}(t)+b(t))]
+\sum_{n=2} \frac{(-1)^n}{n!}[(b^{\dag}(t)+b(t)),S^n])]\\
&= {\rm Tr}[e^{S}(b^{\dag}(t)+b(t)) (-\sum_{n=2}\frac{1}{n!}[S^{(n)},(b^{\dag}(t)+b(t))]
+\sum_{n=2} \frac{(-1)^n}{n!}[(b^{\dag}(t)+b(t)),S^n])]=\frac{1}{2}	{\rm Tr}[2\omega \Delta x(t) ^2] 	.
    			\end{aligned}
\end{equation}
where $S^{(n)}$ denote $n$-th order commutator. Then we have
\begin{equation} 
	\begin{aligned}
	{\rm Tr}[2\omega \Delta x(t) ^2] =-2
 {\rm Tr}[e^{S}(b^{\dag}(t)+b(t)) S[S,(b^{\dag}(t)+b(t))]]	
 +O(S^3).
    			\end{aligned}
\end{equation}

\section{Appendix.B: Haar random unitary in Schrödinger-picture}

\begin{figure}
		\includegraphics[width=0.8\linewidth]{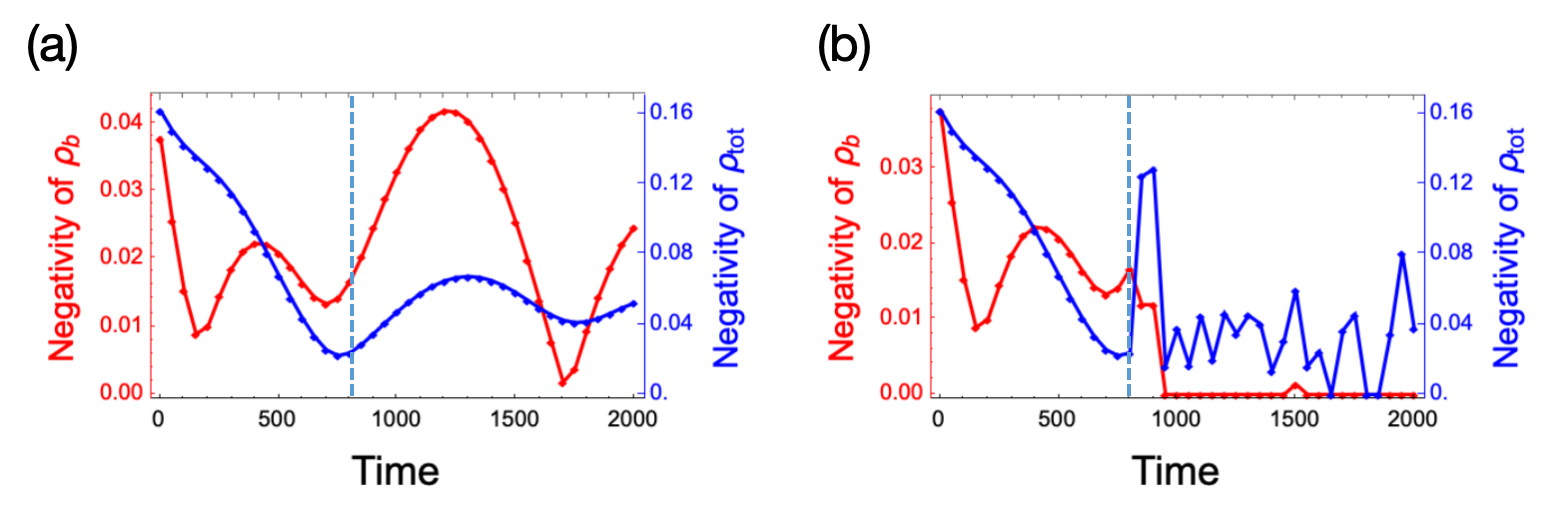}
		\caption{(a) Negativity of $\rho_{b}$ and $\rho_{tot}$ for unitary evolution (infinitely small threshold value).
		(b)
		Negativity of $\rho_{b}$ and $\rho_{tot}$ for unitary-to-nonunitary transition (finite threshold value).
		A peak of negativity (or internal entanglement for $\rho_{tot}$ near $t_{c}$ 
		is consistent with the sudden turn-on of system-environment coupling due to the nonunitary dynamics,
	and also potentially related to the quantum criticality due to the emergent degenerate ground states
	(a highly correlated ground state manifold).
	One can also notice a transition from power-law decay to exponential decay from the negativity of $\rho_{tot}$
before $t_{c}$ (indicated by the vertical dash-line), consistent with the nonMarkovian-to-Markovian transition.	
Since for total density there are three qubits (one of electron subsystem and two of boson subsystem),
 the negativity for any bipartition for maximally entangled state is 1/2 (GHZ state).
Thus here $\rho_{tot}$ is initially partially entangled,}
		\label{nega}
\end{figure}

In terms of the density matrices, when exceeds the critical time $t_{c}$, the degeneracy 
of $H$ causes additional disorder such that the system's evolution is no longer purely unitary. This is equivalent to considering the perturbation itself as inherently random. In this case, scrambling should be observed at the statistical average level as the randomness change the density operator in each run, 
differing from the deterministic unitary evolution at $t<t_{c}$.
Thus in later stage of evolution, the outcomes of measurement on $\rho_{tot}$, which are
separable pure states, consititute Haar ensemble with $\rho_{tot}$ approaches to the maximally mixed state 
(a classical statistical mixture or incoherence superposition of pure states)
with minimal internal entanglement (maximal scrambling with the environment) and accessible information\cite{Chang}.
In this stage, the thermal state (total density) tend to separable at non-zero temperature 
and each subsystem also becomes
higher mixed due to the rised entanglement with environment (see Fig.\ref{f7}(d) and Fig.\ref{nega}(b)).

For separable pure state representing the outcome of measurement on $\rho_{tot}$,
$|\Psi_{i}\rangle\langle\Psi_{i}|$ (the projector which is fixed in time),
the corresponding probability for finding the system in computational basis state $|\Psi_{i}\rangle$
is ${\rm Tr}[\rho_{tot}|\Psi_{i}\rangle\langle\Psi_{i}|]
=|\langle \psi_{0}|\Psi_{i}\rangle|^2$,
which is also the expectation of observable $|\Psi_{i}\rangle\langle\Psi_{i}|$ on the state $\rho_{tot}$.
We define a Haar random unitary 
(maximal randomness due to the maximal system-bath entanglement and thus
maximally mixed on the outcomes from system) $U$ that satisfies
$\rho_{tot}(t)=U \rho_{tot}(0) U^{\dag}$ (with $\rho_{tot}(0)$ be Hermitian) before the $\rho_{tot}$ becomes mixed state.
Then for non-separable states $\rho_{tot}(0)$ and its copy $\sigma_{tot}(0)$,
which are both Hermitian and positive semi-definite,
the global Haar-random unitary satisfies (1-design and 2-design, respectively)
\begin{equation} 
	\begin{aligned}
&\int dU U\rho_{tot}(0)U^{\dag}=\frac{{\bf I}}{N},\\
&\int dU (U\otimes U)(\rho_{tot}(0)\otimes \sigma_{tot}(0))(U^{\dag}\otimes U^{\dag})
=\int dU (U\rho_{tot}(0)U^{\dag})\otimes (U\sigma_{tot}(0))U^{\dag})\\
&
=\frac{ 1-{\rm Tr}[\rho_{tot}(0)\sigma_{tot}(0)]}{N^2-1}{\bf I}\otimes {\bf I}
+\frac{N{\rm Tr}[\rho_{tot}(0)\sigma_{tot}(0)]-1}
{N(N^2-1)}\times swap\ operator\ on\ H_{tot}\otimes H_{tot},
    			\end{aligned}
\end{equation}
with $0\le {\rm Tr}[\rho_{tot}(0)\sigma_{tot}(0)]\le 1$.
Choosing the finite ensemble of unitaries with lowered quantum randomness,
the approximate $k$-design is allowed by replacing the integral over continuous 
$U$ with the summation over the discrete ones.
While the statistical average (instead of the classical one) of the rotated projector in the Heisenberg picture
over $U$ and outcomes reads
\begin{equation} 
	\begin{aligned}
\mathbb{E}_{U,i}[U^{\dag} |\Psi_{i}\rangle\langle\Psi_{i}| U]=
\int dU\sum_{i}{\rm Tr}[U\rho_{tot}(0)U^{\dag}  |\Psi_{i}\rangle\langle\Psi_{i}| ]
U^{\dag} |\Psi_{i}\rangle\langle\Psi_{i}| U
=\int dU\sum_{i}{\rm Tr}[\rho_{tot}(0)U^{\dag} |\Psi_{i}\rangle\langle\Psi_{i}| U]
U^{\dag} |\Psi_{i}\rangle\langle\Psi_{i}| U,
    			\end{aligned}
\end{equation}
which is valid even for nonseparable $\rho_{tot}(0)$.
Applying unitaries from a Haar random ensemble (or a sufficiently high-order k-design) 
to $\rho_{tot}(0)$ is to realize a depolarizing effect,
such that the remaining polarizing effect is due to the internal entanglement in $\rho_{tot}$,
which force the measurement outcomes to follow a correlated distribution as a hallmark of entanglement.
A typical example is the continuous measurement performed on open system 
that is strongly coupled to a non-Markovian environment and exhibit decoherence dynamics\cite{Shabani}.
Only when measurement outcomes are uniformly random, i.e., each with equal probability $1/N$,
the $U\rho_{tot}(0)U^{\dag}$ as well as $\rho_{tot}(0)$ is a maximally mixed state ${\bf I}/N$.

Similar randomness can be extracted in terms of classical shadow protocol 
by the ensemble of measurements, but for seperable initial density\cite{Mok},
i.e., the average over $U$ can be replaced by average over the 
uniformly and randomly chosen pure computational basis states of a subsystem.
$\mathbb{E}_{U,i}[U^{\dag} |\Psi_{i}\rangle\langle\Psi_{i}| U]=\frac{{\bf I}}{N}$ only in the case of unentangled
subsystems $\rho_{e}$ and $\rho_{b}$,
in which case the uniform summation over basis states have $\sum_{i} |\Psi_{i}\rangle\langle\Psi_{i}|={\bf I}$.
In terms of above random outcomes (equal footing),
the initial density $\rho_{tot}(0)$ can be obtained through the inverse (unscrambling) map,
such that $\rho_{tot}(0)
=(N+1)\mathbb{E}_{U,i}[U^{\dag} |\Psi_{i}\rangle\langle\Psi_{i}| U]
-\mathbb{E}_{U,i}[{\rm Tr}[U^{\dag} |\Psi_{i}\rangle\langle\Psi_{i}| U]{\bf I}]$\cite{Hu}.

Large fluctuations emerge since $t_{c}$ as shown in Figs.5-6 is anology to the effect of 
measurements that extract the informations of the global (system and environment) correlation as well as quantum fluctuations from the measurement outcomes of $\rho_{tot}$, and the inherent probabilistic nature of quantum collapse
increases the mixedness of system states.
Consistent with the result of nonrevealing measurement,
the fully scrambled ensemble $\rho_{E}(t=\infty)={\bf I}_{E}/N_{E}$
has minimal information about the initial state's quantum fluctuations (insensitive to microscopic details) but maximal information about the global correlations (such that the information can be extracted via measurements on $\rho_{tot}$). 
To accessing this propose, the projected ensemble $\rho_{E}$
is expressed as a classical statistical ensemble (probability distribution over the replicated
Hilbert space) of a set of separable pure states\cite{Chang}.
So compares to $\rho_{E}(t>t_{c})$, $\rho_{E}(t\rightarrow\infty)$ has a converged form and reflecting the
statistical distribution of states in Haar ensemble and thus has lowerest purity.

For $t<t_{c}$, one observes a monotonous decay of negativity due to the unitary scrambling
which indicates a transition from power-law decay to exponential decay,
while for $t>t_{c}$, the stronger fluctuation in internal entanglement
(as can be seen from the negativities in Fig.\ref{nega}) indicates the emergence of non-Markovian dynamics
(due to the degeneracy of $H$ as well as the non-unitary dynamics), which obscures
the Markovian dynamics at long-time.
This is inevitable as long as it is related to the density operator which is
inherently dependent on the total Hamiltonian, like the expectation ${\rm Tr}[x_{1}(t)\rho_{tot}(t)]
={\rm Tr}[x_{1}(t)|\psi_{0}\rangle\langle\psi_{0}|]$,
with $|\psi_{0}\rangle=\sum_{i} \langle\Psi_{i}|\psi_{0}\rangle |\Psi_{i}\rangle$.

Overall,
the $\rho_{tot}$ remains nonseparable throughout the evolution, from pure state to mixed state.
The subsystems $\rho_{e}$ and $\rho_{b}$ are entangled throughout the evolution,
where $\rho_{e}$ (single-qubit system with classical correlations) remains separable throughout the evolution,
while $\rho_{b}$ is nonseperable initially but the strength of (internal) entanglement decreases over time, 
and becomes seperable at the moment when the degenerated ground state of $H$ appears
(which can be verified by calculating the negativity). Thus, time evolution primarily leads to two changes: the emergence of entanglement between the total system and its external environment (resulting in information loss), and the diminution of internal entanglement within the phonon subsystem.

\end{document}